\documentclass[aps,prl,twocolumn,superscriptaddress,showpacs]{revtex4}
\usepackage{hyperref}

\usepackage{graphicx}

\usepackage{amsmath}
\usepackage{epsfig}

\begin{document}
\title{Experimental demonstration of coherent state estimation with minimal disturbance}

\author{Ulrik L. Andersen}
\affiliation{Institut f\"{u}r Optik, Information und Photonik,  Max-Planck Forschungsgruppe, Universit\"{a}t Erlangen-N\"{u}rnberg, G\"{u}nther-Scharowsky str. 1, 91058, Erlangen, Germany}
\email{andersen@kerr.physik.uni-erlangen.de}
\author{Metin Sabuncu}
\affiliation{Institut f\"{u}r Optik, Information und Photonik,  Max-Planck Forschungsgruppe, Universit\"{a}t Erlangen-N\"{u}rnberg, G\"{u}nther-Scharowsky str. 1, 91058, Erlangen, Germany}
\author{Radim Filip}
\affiliation{Institut f\"{u}r Optik, Information und Photonik,  Max-Planck Forschungsgruppe, Universit\"{a}t Erlangen-N\"{u}rnberg, G\"{u}nther-Scharowsky str. 1, 91058, Erlangen, Germany}
\affiliation{Department of Optics, Research Center for Optics, Palacky University, 17. Listopadu 50, 77200 Olomouc
Czech Republic}
\author{Gerd Leuchs}
\affiliation{Institut f\"{u}r Optik, Information und Photonik,  Max-Planck Forschungsgruppe, Universit\"{a}t Erlangen-N\"{u}rnberg, G\"{u}nther-Scharowsky str. 1, 91058, Erlangen, Germany}

\date{\today}

\begin{abstract}
We investigate the optimal tradeoff between information gained about an unknown coherent state and the state disturbance caused by the measurement process. We propose several optical schemes that can enable this task, and we implement one of them, a scheme which relies on only linear optics and homodyne detection. Experimentally we reach near optimal performance, limited only by detection inefficiencies. In addition we show that such a scheme can be used to enhance the transmission fidelity of a class of noisy channels.
\end{abstract}


\maketitle

An observer who receives an unknown quantum state, from a set of nonorthogonal states, cannot perfectly retrodicts which state he got. 
Furthermore in the act of trying to reveal the identity of the quantum state, the observer will inevitably alter the state~\cite{bennett92.prl,ekert94.pra}. Obviously there is a tradeoff between maximal extraction of information and minimal disturbance of the state~\cite{fuchs96.pra,banaszek01.prl,fuchs01.pra,banaszek01.pra,barnum02.xxx}. E.g. if the observer applies the best possible measurement strategy allowed by quantum mechanics by which he gains maximal information, the state will be maximally disturbed and vice versa. The fundamental and intriguing features of this tradeoff have fuelled an explosion of research since the early days of quantum mechanics. Recently, however this research has become of practical relevance since it underpins the security of quantum key distribution schemes~\cite{crypto}. 

The balance between information gain and disturbance has hitherto mainly been studied in finite dimensional systems, where inequalities stating the optimal tradeoff have been established for various cases~\cite{banaszek01.prl,banaszek01.pra,mista05.pra} and demonstrated recently in an experiment ~\cite{sciarrino05.xxx}. 
In contrast, very little work has been devoted to the study of this tradeoff in infinitely dimensional systems where quantum information is carried by observables with a continuous spectrum~\cite{ralph00.pra}, important examples being the canonically conjugate quadrature amplitudes.  
A particular class of continuous variable states which have played a key role in various experimental realizations of quantum information protocols is the class of Gaussian states, since they are experimentally easy to produce and manipulate~\cite{braunstein05.rev}. 

In this Letter we investigate the tradeoff between information gain and state disturbance for completely unknown coherent states. Under the assumption that the Gaussian statistic must be preserved we derive the optimal tradeoff, stated in terms of an inequality using appropriate measures for the information gain and state disturbance. The optimal tradeoff can be implemented experimentally using a remarkable simple setup requiring only linear optics and homodyne detection. Besides being fundamentally interesting, this scheme is capable of increasing the transmission fidelity of some noisy channels, an improvement that will be also demonstrated in this Letter.    

Completely unknown pure quantum states described in an infinitely dimensional Hilbert space can be estimated only very poorly  based on a single measurement. However, in quantum communication systems, a priori knowledge is often given. It is e.g. normally known that the state belongs to a certain set each occurring with a certain a priori probability. In this work we assume that the states are taken from a flat distribution of coherent states. With this a priori information at hand, it was recently proven that the state can be optimally estimated using a setup where conjugate quadratures are measured simultaneously using a symmetric beam splitter and two homodyne detectors~\cite{hammerer05.prl}. However, employing this strategy the coherent state is maximally disturbed. On the contrary, if the unknown coherent state is left untouched, our guess will be completely random but the state will be intact. In the following we will investigate the intermediate cases and hence address the question: For a given information gain what is the minimum disturbance to the coherent state? 

Consider a coherent state characterised by the amplitude and phase quadrature $\hat{x}_{in}$ and $\hat{p}_{in}$ with $[\hat{x}_{in},\hat{p}_{in}]=2i$. The state is injected into a machine with a classical output and a quantum output. In general the function of such a machine, under the assumption that the Gaussian statistics is preserved, can be described by a generic linear transformation:
\begin{eqnarray}
\hat{x}_{out}&=&g(\hat{x}_{in}+\hat{n}_x) \,\,\,\,\,\,\, \hat{p}_{out}=g(\hat{p}_{in}+\hat{n}_p)\\
\hat{x}_m &=& \hat{x}_{in}+\hat{m}_x \,\,\,\,\,\,\,\, \hat{p}_m = \hat{p}_{in}+\hat{m}_p\nonumber
\label{transformation}
\end{eqnarray}
where $\hat{x}_{out}$ and $\hat{p}_{out}$ are the amplitude and phase quadrature operators of the quantum state after the interaction, and $\hat{x}_m$ and $\hat{p}_m$ are the directly measured operators (and thus commuting). $g$ is the gain of the operation, $\hat{n}_x$ and $\hat{n}_p$ are operators associated with the noise added to conjugate quadratures of the quantum state, and $\hat m_x$ and $\hat m_p$ are operators responsible for the added noise in the measurement. Now using the commutation relations for the input and output quadratures and assuming that $\hat{x}_{in}$ and $\hat{p}_{in}$ commutes with the noise operators we easily find the following commutation relations:
\begin{eqnarray}
&&[\hat{n}_x,\hat{n}_p] = 2i\frac{1-g^2}{g^2}\label{comm1}\\ 
&&[\hat{m}_x,\hat{m}_p] = [\hat{n}_x,\hat{m}_p]=[\hat{m}_x,\hat{n}_p]=-2i \label{comm2}
\end{eqnarray}
Assuming phase insensitive operation (meaning that conjugate quadratures are equally well measured and equally disturbed), we derive the tradeoff relation for the added noises from the commutation relations to be~\cite{ralph00.pra}
\begin{equation}
\Delta^2 \hat{n} \Delta^2 \hat{m} \geq 1
\label{tradeoff1}
\end{equation}
with the constraints $\Delta^2 \hat{m} \geq 1$ and $\Delta^2 \hat{n}\geq (|1-g^2|)/g^2$. 
$\Delta^2 \hat{n}$ denotes the uncertainty with which the coherent state can be estimated and $\Delta^2 \hat{m}$ is the variance of the noise added to the quantum state caused by the measurement. Therefore the inequality in (\ref{tradeoff1}) dictates that the optimum tradeoff for the added noises is achieved for $\Delta^2 \hat{n} \Delta^2 \hat{m} = 1$. However, this relation can only be satisfied when the gain $g$ can be chosen freely to minimize the added noises. Universality of the protocol, however, requires unity gain operation ($g=1$) which imposes another restriction to the achievable tradeoff:
By noting that the commutation relations for the added noise to the measurement in (\ref{comm2}) resembles that of quadrature operators, we set $\hat{m}_x=\hat{x}_1$ and $\hat{m}_p=-\hat{p}_1$ where $\hat{x}_1$ and $\hat{p}_1$ are quadratures of the ancilla. The commutation relation (\ref{comm1}) then forces the remaining noise operators into the superpositions $\hat{n}_x=\hat{x}_1+\hat{x}_2$ and $\hat{n}_p=\hat{p}_1-\hat{p}_2$, where $\hat{x}_2$ and $\hat{p}_2$ likewise are quadratures of the other ancilla. The task is now to simultaneously minimise the variance of $\hat{x}_1+\hat{x}_2$ and $\hat{p}_1-\hat{p}_2$ for given variances of $\hat{x}_1$ and $\hat{p}_1$. This can be done optimaly by using a two-mode squeezed ancilla, and for the phase insensitive case we find the following tradeoff relation: $\Delta^2 \hat{n}\geq 2(\Delta^2 \hat{m}-\sqrt{(\Delta^2 \hat{m})^2-1})$.

For quantification of the information gain, we use the estimation fidelity $G$, which is the phase space overlap between the input state and the state that can be prepared based on the classical information and the a priori information~\cite{furusawa98.sci,hammerer05.prl}. The disturbance is quantified by the transfer fidelity $F$, which quotes the overlap between the input state and the post measurement state~\cite{furusawa98.sci,hammerer05.prl}. With unity gain the fidelities are simply given by $G=2/(3+\Delta^2 \hat{m})$ and $F=2/(2+\Delta^2 \hat{n})$. 
In terms of the fidelities the tradeoff relation therefore reads~\cite{note}
\begin{equation}
F\leq \frac{G}{2\left( 1-G-\sqrt{(1-G)(1-2G)}\right)}
\label{tradeoff2}
\end{equation}

\begin{figure}[h] \centering \includegraphics[width=6cm]{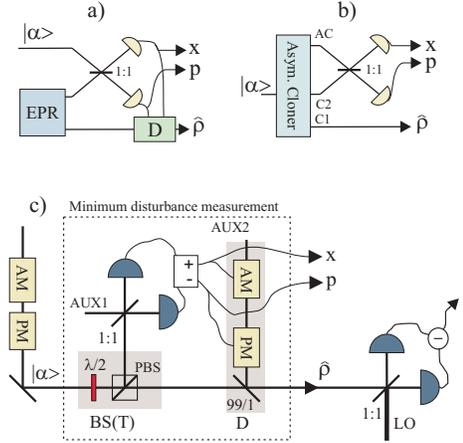} \caption{\it Three methods by which the optimal tradeoff can be established. a) Teleportation. b) Asymmetric cloning followed by a joint measurement. c) Simple feed forward approach. C(1,2): Clones,  AC: Anti-clone, D:Displacer, AM: Amplitude modulator, PM: Phase modulator, PBS: Polarizing beam splitter, BS(T): Variable beam splitter with transmission T, AUX(1,2): Auxiliary beams and LO: Local oscillator.}
\label{setup}  \end{figure}

The optimal tradeoff between classical and quantum information can be implemented using various systems. One strategy is to use teleportation~\cite{furusawa98.sci}, where the measurement outcomes at the Bell state analyzer yield the classical information whereas the teleported state serves as the quantum state after disturbance (see Fig. \ref{setup}a). The whole range of optimal tradeoffs can then be achieved by tuning the amount of entanglement. Alternatively, a continuous variable asymmetric quantum cloner~\cite{fiurasek01.prl} can be employed as illustrated in Fig. \ref{setup}b: The anti-clone is mixed with one of the clones on a symmetric beam splitter, and subsequently amplitude and phase quadratures are measured in the two output ports to retrieve some classical information. The other clone is left unchanged and serves as the output quantum state. By tuning the asymmetry of the cloning machine, and the beam splitting ratio of the measuring beam splitter accordingly, the whole range of maximal tradeoffs can be accessed. 
 
Interestingly, a much simpler approach can realise the optimal tradeoff. The scheme, which relies entirely on simple linear optical components and homodyne detectors is depicted inside the dashed box of Fig. \ref{setup}c: The quantum state is partially reflected off a beam splitter with transmission coefficient $T$. The reflected part of the state is optimally estimated by simultaneously measuring two conjugate quadratures, i.e. $\hat x$ and $\hat p$, and the resulting classical information is partly used to guess the state and partly used to displace the transmitted part of the quantum state after an appropriate scaling of the classical data. Using this very simple scheme, the added noises to the quantum state are $\Delta^2 \hat{n}=2(1-\sqrt{T})^2/(1-T)$ and $\Delta^2 \hat{m}=(1+T)/(1-T)$ and the estimation and transfer fidelities are found to be
\begin{eqnarray}
G=\frac{1-T}{2-T}\,\,\,\,\,\,\,\,\,F=\frac{1-T}{2-2\sqrt{T}}
\end{eqnarray}
which is saturating inequality (\ref{tradeoff2}). 

A distinct difference between the first two approaches in Fig. \ref{setup}a and \ref{setup}b (teleportation and cloning) and the simple approach in Fig. \ref{setup}c is that the latter one does not require any nonlinear interaction. Some similarities between the teleportation scheme and the feed forward scheme were pointed out in ref. \cite{hofmann01.pra}, and both schemes have been suggested as potential eavesdropping attacks~\cite{ralph00.pra,lance04.prl}. However we note that the cloning approach in Fig.~\ref{setup}b might be superior for an eavesdropper, since in this protocol the classical information can be extracted at any instance if a quantum memory is available.     

We now proceed to the experimental demonstration of the optimal tradeoff using the simple approach. The setup (which is schematically shown in Fig. \ref{setup}c) consists basically of three stages: A preparation stage where the coherent state is prepared, a separation stage in which classical information is separated from the quantum information and finally a verification stage where the quantum state is characterised. 
To ensure high purity of the input state we define our states to reside at a certain sideband frequency within a certain spectral window. With this definition an arbitrary coherent state can be easily generated by modulating the laser beam with an amplitude (AM) and a phase modulator (PM). We chose the sideband to have a bandwidth of 100~kHz with a center frequency of 14.3~MHz, since at this frequency the laser (Nd:YAG oscillating at 1064nm) was quantum noise limited. 

After the preparation stage, the coherent state is injected into the separation stage. Here the state is divided into two parts by the variable beam splitter (BS(T)), which is composed of a half wave plate and a polarizing beam splitter; any transmission is therefore easily accessed by a simple wave plate rotation. The reflected part is estimated by performing simultaneous measurements on the amplitude and phase quadratures as shown in Fig.~\ref{setup}c: The signal interferes at a beam splitter with an auxiliary beam (AUX1) with a $\pi$/2 relative phase shift and balanced intensities, and subsequently the two outputs are measured and the difference and sum currents are generated. These two outputs then provide information about the phase and amplitude quadratures of the signal, which is then used to displace the remaining quantum state in order to ensure unity gain operation. This is done by modulating an auxiliary beam (AUX2) using an amplitude and a phase modulator, and subsequently combine this beam at a 99/1 beam splitter with the remaining signal~\cite{furusawa98.sci}.   
Finally, after the information retrieval and displacement, the resulting quantum state is characterised using a standard homodyne detector with a strong local oscillator (LO). The signal and the noise variances of the phase and the amplitude quadratures are measured using a spectrum analyzer with resolution and video bandwidth set to 100~kHz and 30~Hz, respectively. These variances fully characterize the output state. We compute the transfer fidelity by comparing the output to the input state, which was measured using the same homodyne detector in order to make a consistent comparison~\cite{furusawa98.sci}. The estimation fidelity is calculated from the carefully measured reflectivity of the variable beam splitter. In order to avoid erroneously overestimations of the fidelities the values are corrected to account for detection inefficiencies. 

\begin{figure}[h] \centering \includegraphics[width=7cm]{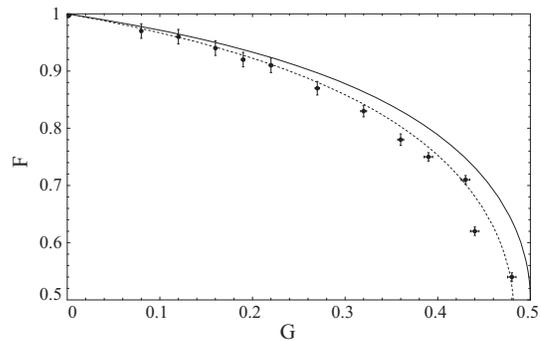} \caption{\it Quantum state fidelity as a function of the estimation fidelity. The optimal tradeoff given by Eq. \ref{tradeoff2} is represented by the solid curve, whereas the dashed curve is associated with the tradeoff taking into account detection inefficiencies. The error bars stem from the inaccuracy in determining the detector efficiencies.}
\label{fidelity}  \end{figure}

Many different tradeoffs were realised and the results are shown in Fig.~\ref{fidelity}.  
The solid curve in Fig.~\ref{fidelity} represents the optimal tradeoff given by the saturation of inequality (\ref{tradeoff2}). Deviation from optimality is caused by the inefficiency of the in-loop detector which partly degrades the classical guess and partly imposes additional non-fundamental noise onto the quantum state after measurement. Therefore we paid special attention to the optimization of this measurement: The mode matching efficiency between the auxiliary beam (AUX1) and the signal state was carefully optimized to yield a visibility of 99\% and the quantum efficiency of the photo diodes were 95\%. Despite the high quality of the homodyne setup, the curve for the optimum tradeoff achievable with the experimental setup is slightly shifted and represented by the dashed curve in Fig.~\ref{fidelity}.     

\begin{figure}[h] \centering \includegraphics[width=7cm]{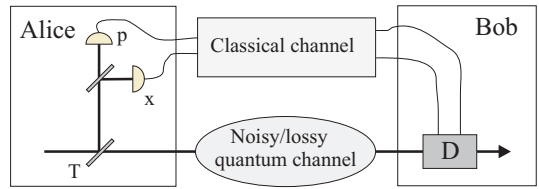} \caption{\it Schematic illustration of a protocol capable of increasing the fidelity of noisy channels. The displacement operation, D, is similar to the one shown in the long shaded box of Fig.~\ref{setup}c, and other experimental details about the preparation and verification follow those of Fig.~\ref{setup}c.}
\label{application}  
\end{figure}

In the last part of the paper we discuss how optimal partial state estimation can be exploited to enhance some communication tasks.   
Let us consider the following protocol. Alice wants to transmit quantum information which is encoded into a coherent state, and after the transmission Bob receives the quantum state. Such a communication task is always inflicted by loss or noise, hereby corrupting the quantum state and as a result reducing the transmission fidelity. Let us first consider a lossy channel characterized by the transmission coefficient $\eta$. In such a channel the transmission loss must be compensated by an amplifier in order to ensure maximal transmission fidelity. The highest transmission fidelity is achieved by amplifying the state before it is injected into the lossy channel. However, by considering the power constraint scenario~\cite{holevo}, amplification prior to transmission is not possible and as a result the amplifier must be placed at Bob's receiving station. This yields a transmission fidelity of $F=\eta$. However, by optimally separating the input state into a classical and a quantum channel as demonstrated in this article, the fidelity can be increased: The optimal post measurement quantum state is sent through the lossy channel, whereas the classical information is sent through a classical channel (see Fig.~\ref{application}). At the receiving station Bob displaces the corrupted quantum state based on the information he gains from the classical channel. Obviously there is an optimal separation ratio between the classical and the quantum information for a given attenuation in the channel. This ratio is optimized by setting $T=\eta$ and we find the optimised fidelity to be $F=1/(2-\eta)$. This fidelity is for all values of $\eta$ larger than the fidelity achievable when only the quantum channel is used. 

We demonstrate this idea by inserting an attenuator with $\eta=31\%$ into the channel, which is placed between the variable beam splitter (BS(T)) and the displacement operation (D) in Fig.~\ref{setup}c (or Fig.~\ref{application}). If an amplifier is employed after the channel to compensate for these losses the fidelity is $F=31\%$. Now using our strategy of dividing the information into classical and quantum as shown in Fig.~\ref{application}, we measure a quantum state fidelity of $F=63\pm 1\%$, which clearly surpasses the standard amplifier approach. In this experimental run we measured the gains to be $1.00\pm0.01$ and $1.01\pm0.01$ for the amplitude and phase respectively. 

We now consider a fully transparent channel which adds noise to the signal. We first assume that the nature of this noise is additive and deterministic. If the added noise of the quantum channel exceeds two vacuum units, pure classical communication maximises the fidelity and is therefore the better alternative. However if the added noise is less than two vacuum units, then pure quantum communication becomes advantageous.  Therefore, only the two extreme schemes will be relevant. From here on we assume that the noise in the channel is additive and probabilistic. In this case  the intermediate scheme also becomes important~\cite{ricci05.prl}: We consider a channel which is perfectly transmitting the signal with probability $p$ and fails to transmit it with probability $(1-p)$. The average fidelity of such a channel is given
by $F=p$. If we now apply the partial estimation approach in front of the channel as illustrated in Fig.~\ref{application}, then the fidelity is given by $F'=pF+(1-p)G$.
We find that for $0<p<4/5$, optimal partial estimation increases the transmission fidelity. If on the other hand $p\ge 4/5$, the best strategy is to entirely use the quantum channel. The fidelity is maximized only for a specific separation of the quantum and classical information which depends on the probability $p$. As $p$ approaches zero it is best to estimate completely the signal and send it through the classical channel, and as $p$ approaches $4/5$, it is best to send the entire signal through the quantum channel. For all intermediate cases the optimal partial estimation is advantageous. The optimal solution, which is found by solving an algebraic equation, is computed numerically, and we find that the maximal improvement appears for $p=0.5$ where the fidelity improves by approximately 10\% if $T=0.405$.

In this Letter we have extended the discussion on the optimal information-disturbance tradeoff to the continuous variable regime and derived a tradeoff relation for coherent states. Furthermore, we have proposed a simple linear optics circuit which saturates this relation, and we have implemented the idea and obtained near optimal performance. Finally, we have demonstrated that our scheme can be used to enhance the transmission fidelity of some noisy channels, rendering our approach as a useful tool in future quantum communication networks.

We thank Ladislav Mista and Vincent Josse for fruitful discussions. This work has been supported by the EU projects COVAQIAL (project no. FP6-511004). R.F. was supported by the projects: 202/03/D239 of GACR, MSM6198959213 of
MSMT CR and by the Alexander von Humboldt foundation.


\begin{thebibliography}{99}

\bibitem{bennett92.prl} C.~H.~Bennett et al., Phys. Rev. Lett. {\bf 68}, 557 (1992).
\bibitem{ekert94.pra} A.~K.~Ekert et al., Phys. Rev. A {\bf 50}, 1047 (1994). 
\bibitem{fuchs96.pra} C. A. Fuchs, and A. Peres, Phys. Rev. A {\bf 53}, 2038 (1996).
\bibitem{banaszek01.prl} K.~Banaszek, Phys. Rev. Lett. {\bf 86}, 1366 (2001).
\bibitem{fuchs01.pra} C. A. Fuchs, and K. Jacob, Phys. Rev. A {\bf 63}, 062305 (2001).
\bibitem{banaszek01.pra} K. Banaszek, and I. Devetak, Phys. Rev. A {\bf 64}, 052307 (2001).
\bibitem{barnum02.xxx} H. Barnum, e-print quant-ph/0205155
\bibitem{crypto} N. Gisin et al, Rev. Mod. Phys. {\bf 74}, 145 (2002).
\bibitem{sciarrino05.xxx} F. Sciarrino et al., e-print quant-ph/0510097. 
\bibitem{mista05.pra} L. Mista Jr. et al., Phys. Rev. A {\bf 72}, 012311 (2005).
\bibitem{ralph00.pra} T.~Ralph, Phys. Rev. A {\bf 62}, 62306 (2000).
\bibitem{braunstein05.rev} S. L. Braunstein and P.v.Loock, Rev. Mod. Phys. {\bf 77}, 513 (2005).
\bibitem{hammerer05.prl} K. Hammerer et al, Phys. Rev. Lett. {\bf 94}, 150503 (2005).
\bibitem{note} The derivation of inequality (\ref{tradeoff2}) is based on the constraint that the Wigner function of the quantum state remains Gaussian. After the completion of this work, it was realised that if one allows for non-Gaussian features of the output state, the fidelity can be slightly improved using a clever (but experimentally very challenging) input ancilla state. L. Mista Jr. e-print quant-ph/0510191.
\bibitem{furusawa98.sci} A. Furusawa et al. Science {\bf 282}, 706 (1998).
\bibitem{fiurasek01.prl} J. Fiurasek, Phys. Rev. Lett. {\bf 86}, 4942 (2001).
\bibitem{hofmann01.pra} H.~F.~Hofmann et al., Phys. Rev. A {\bf 64}, 040301 (2001).
\bibitem{lance04.prl} C. Weedbrook et al., Phys. Rev. Lett. {\bf 93}, 170504 (2004).
\bibitem{holevo} C.M Caves and P. Drummond, Rev. Mod. Phys. {\bf 66}, 481 (1994). 
\bibitem{ricci05.prl} M.~Ricci et al., Phys. Rev. Lett. {\bf 95}, 90504 (2005).


\end{thebibliography}
\end{document}